\documentclass[prd,superscriptsize,onecolumn,nofootinbib,notitlepage]{revtex4-1}
\usepackage[utf8]{inputenc} 

\usepackage{graphicx,xcolor,overpic,mathtools}
\usepackage{amsthm,amsmath,amssymb}
\usepackage[colorlinks]{hyperref}
\definecolor{coolblack}{rgb}{0.0, 0.18, 0.39}
\hypersetup{
    colorlinks = true,
    citecolor = {coolblack},  
    urlcolor = {blue}
   }
\PassOptionsToPackage{normalem}{ulem}
\usepackage{ulem} 
\usepackage{sidenotes}
\usepackage{bm}
\usepackage{url}
\usepackage{physics}
\usepackage[makeroom]{cancel}
\usepackage{tensor}
\usepackage{mathrsfs}
\usepackage{slashed}
\usepackage{tikz}
\usepackage[mode=buildnew]{standalone}
\usepackage{subcaption}

\newcommand{\fpi}{f_{\pi}}
\newcommand{\mpi}{m_{\pi}}
\newcommand{\lag}{\mathcal{L}}

\newcommand{\comment}[1]{}

\NewDocumentCommand{\evat}{sO{\bigg}mm}{%
  \IfBooleanTF{#1}
   {\mleft. #3 \mright|_{#4}}
   {#3#2|_{#4}}%
}
\usepackage{enumerate}
\usepackage{cleveref}

\usepackage{stackengine}
\stackMath

\DeclareMathOperator{\Mev}{MeV}
\DeclareMathOperator{\fm}{fm}

\begin{document}
\title[
]{Gravitational form factors of nuclei in the Skyrme model}

\author{Alberto García Martín-Caro}
\email{alberto.martin-caro@usc.es}
\author{Miguel Huidobro}
\email{miguel.huidobro.garcia@usc.es}
\affiliation{Departamento de F\'isica de Part\'iculas, Universidad de Santiago de Compostela and Instituto
Galego de F\'isica de Altas Enerxias (IGFAE) E-15782 Santiago de Compostela, Spain}
 \author{Yoshitaka Hatta}
 \email{yhatta@bnl.gov}
\affiliation{Physics Department, Brookhaven National Laboratory, Upton, NY 11973, USA}
\affiliation{RIKEN BNL Research Center, Brookhaven National Laboratory, Upton, NY 11973, USA}

\date[ Date: ]{\today}
\begin{abstract}
We compute the gravitational form factor $D(t)$ of various nuclei in the generalized Skyrme model where  nuclei are described as  solitonic field configurations each with a definite baryon number $B$. We separately discuss  the cases $B=1$ (nucleons), $B=2$ (deuteron), $B=3$ (helium-3 and tritium) and extrapolate to larger $B$-values. 
Configurations with $B>1$  are in general not spherically symmetric, and we demonstrate how group theory helps to extract the form factor. 
Numerical results are presented for the configurations with $B=1,2,3,4,5,6,7,8,32,108$. We find that the $B$-dependence is consistent with a power-law $D(0)\propto B^{\beta}$ with $\beta=1.7\sim 1.8$. 
Other gravitational form factors can be calculated in the same framework, and we show the result for the $J(t)$ form factor associated with angular momentum  for the $B=3$ solution.   

\end{abstract}
\maketitle
\comment{
\begin{minipage}{\textwidth}
\tableofcontents
\end{minipage}
}
\section{Introduction}

The gravitational form factors (GFFs) are  form factors  associated with the QCD energy momentum tensor $T_{\mu\nu}$ \cite{Kobzarev:1962wt,Pagels:1966zza}. Much like the electromagnetic form factors, they contain a wealth of information about the structure of the nucleons, or more generally hadrons and nuclei. However, unlike  the latter which have enjoyed more than 70 years of continual theoretical and experimental  efforts, it was not until recently that GFFs started to attract community-wide attention \cite{Polyakov:2018zvc}. A major catalyst is the Electron-Ion Collider (EIC) project \cite{AbdulKhalek:2021gbh}  poised to uncover the mass and spin structure of the nucleons and nuclei. Since the energy momentum tensor encodes mechanical properties such as mass, angular momentum, internal forces and their distribution, the study of  GFFs is perfectly aligned with the core missions of the EIC. In a sense, GFFs are  more `quintessentially QCD'  than the electromagnetic form factors   since they can be defined without reference to quark electric charges, and are directly sensitive to the gluonic degrees of freedom. 

Among the various GFFs, of particular interest is the $D(t)$ form factor related to the `pressure' distribution inside hadrons and nuclei  \cite{Polyakov:2018zvc}. Its value at vanishing momentum transfer $D(t=0)$, often referred to as the D-term, is a fundamental constant similar to the magnetic moment. While the value is presently unknown, even for the nucleons, in principle the form factor $D(t)$  can be experimentally accessed in Deeply Virtual Compton Scattering  \cite{Teryaev:2005uj,Anikin:2007yh} and near-threshold quarkonium photo- and electro-production in electron-nucleon scattering
\cite{Hatta:2018ina,Boussarie:2020vmu,Hatta:2021can,Sun:2021gmi,Guo:2021ibg,Kharzeev:2021qkd,Mamo:2022eui}. See  \cite{Burkert:2018bqq,Duran:2022xag,Adhikari:2023fcr} for recent experimental efforts.  The same methods could be used to measure the $D(t)$ form factor of light nuclei at the EIC.  

In this paper, we compute  $D(t)$ for a number of nuclei in the Skyrme model \cite{Skyrme:1961vq}. In this model, the nucleon is realized as a finite-energy solitonic configuration of meson fields with a nontrivial topological number identified with the baryon number $B=1$. The model has been successful in explaining the low energy properties of the nucleon \cite{Adkins:1983ya,Schechter:1999hg}.  Subsequently it has been extended to describe light nuclei, such as the deuteron ($B=2$) \cite{Verbaarschot:1986qi,Braaten:1988cc}, the helium-3 or the tritium ($B=3$) \cite{Carson:1990yv} and so on. At the moment, solutions up to $B=108$ are known  \cite{Feist:2012ps,Gudnason:2022jkn} and they have been even extrapolated to infinite crystals ($B=\infty$) \cite{Klebanov:1985qi,Castillejo:1989hq} to discuss the properties of nuclear matter \cite{Adam:2022aes}  and neutron stars \cite{Adam:2020yfv,Lee:2021hrw}. Moreover, low-lying excited states of various nuclei have been   systematically studied \cite{Walhout:1992gr,Manko:2007pr,BjarkeGudnason:2018bju}. Concerning form factors, the electromagnetic and axial form factors have  been calculated for the $B=2$  \cite{Braaten:1988bn} and $B=3$  \cite{Carson:1991fu} solutions,  as well as for larger-$B$ solutions with zero isospin \cite{Karliner:2015qoa}.   However, the application of the  model to  GFFs has been so far limited to the $B=1$ sector (nucleons and their excited states) \cite{Cebulla:2007ei,Perevalova:2016dln,Kim:2020lrs} which is relatively tractable due to the  spherical symmetry of the classical solution. In fact, all the known Skyrmion solutions with $B>1$ are not spherically symmetric. Yet, each solution possesses a (discrete) symmetry group, and techniques from group theory can greatly facilitate the extraction of form factors as demonstrated by Carson \cite{Carson:1991fu} in the context of the electromagnetic form factors. 
We shall adapt this method to the computation of GFFs using the state-of-the-art numerical   Skyrmion solutions.         

It should be mentioned that the form factor $D(t)$ for nuclei has been evaluated in the liquid drop model  \cite{Polyakov:2002yz}, the Walecka model \cite{Guzey:2005ba} and from the nonrelativistic nuclear spectral functions  \cite{Liuti:2005qj}. We shall study the dependence of the D-term on the baryon number and compare the result with these previous works.

\section{Skyrmions in the generalized Skyrme model}

The generalized Skyrme model that we will consider is given by the following Lagrangian density,
\begin{align}
    \mathcal{L}_{\rm SK}=
    -\frac{\fpi^2}{16}\Tr\{L_{\mu}L^{\mu}\} + \frac{1}{32e^2}\Tr\{[L_{\mu}, L_{\nu}]^2\} - \lambda^2\pi^4B_{\mu}B^{\mu} + \frac{\mpi^2\fpi^2}{8}\Tr\{U - \bm{1}\},
    \label{Lagrangian}
\end{align}
where we use the metric convention $\eta^{\mu\nu}={\rm diag}(1,-1,-1,-1)$. $U\in SU(2)$ contains the fundamental mesonic degrees of freedom, parametrized as the coordinates of the $SU(2)$ group element
\begin{equation}
    U=\sigma \bm{1}+i\pi_a\tau_a\equiv i\phi^\alpha\bar\tau_\alpha,\quad \phi^\alpha =(\sigma,\bm{\pi}),\quad \Bar{\tau}_\alpha=(-i\bm{1},\bm{\tau}),
\end{equation}
with $\tau^a$ ($a=1,2,3$) being the Pauli matrices. 
$L_{\mu} = U^{\dagger}\partial_\mu U$ are the components of the associated left-invariant Maurer-Cartan form, $\bm{1}$ is the $2\times2$ identity matrix, and $B^{\mu}$ is the topological current,
\begin{equation}
    B^{\mu} = \frac{\epsilon^{\mu\nu\rho\sigma}}{24\pi^2}\Tr\{L_{\nu}L_{\rho}L_{\sigma}\}.
\end{equation}

Although only the first two terms (namely, the quadratic and quartic terms in $L_\mu$) in \eqref{Lagrangian} were originally considered by Skyrme \cite{Skyrme:1961vq}, the other two terms have been subsequently included in order to achieve a better agreement with nuclear phenomenology. Indeed, the potential term explicitly breaks chiral symmetry and provides a mass term for the pions. On the other hand, the sextic term in $L_\mu$, first proposed in \cite{Jackson:1985yz} (see also \cite{Adkins:1983nw}), can be seen as an effective  point-like interaction that describes the repulsive exchange of omega vector mesons, which becomes relevant at sufficiently high densities. Such a term has recently proven to be crucial for an accurate description of the high density equation of state of neutron stars, allowing to reach sufficiently high maximum masses and sound velocities above the conformal limit \cite{Adam:2020yfv, Adam:2022cbs}. 

We will consider static solutions of the Skyrme field, and for numerical purposes we adopt the usual Skyrme units of energy and length,
\begin{equation}
    E_s = \frac{3\pi^2 \fpi}{e} , \quad x_s = \frac{1}{\fpi e},
\end{equation}
unless otherwise specified.
The static energy functional in these units becomes
\begin{align}
    \notag E = \frac{1}{24\pi^2}\int d^3x &\left[ -\frac{1}{2}\Tr\left\{L_i^2\right\} - \frac{1}{4}\Tr\left\{\left[L_i,L_j\right]^2\right\} + 8\lambda^2 \pi^4 f^2_{\pi}e^4 ({B}^0)^2 + \left(\frac{\mpi}{\fpi e}\right)^2 \Tr\{\bm{1} - U\} \right] \\[2mm]
    =\frac{1}{24\pi^2}\int d^3x &\left[ (\partial_i \phi_\alpha)^2 + \left( \partial_i \phi_\alpha \partial_j \phi_\beta - \partial_i \phi_\beta\partial_j \phi_\alpha \right)^2 + c_6\left(\epsilon_{\alpha\beta\gamma\delta}\phi_\alpha\partial_1 \phi_\beta\partial_2 \phi_\gamma \partial_3 \phi_\delta \right)^2 + c_0(1-\sigma) \right],
    \label{Energy_integral}
\end{align}
where $\epsilon_{\alpha\beta\gamma\delta}$  is the totally antisymmetric tensor with $\epsilon_{1234}=1$ and we have defined $c_6 = 2\lambda^2\fpi^2 e^4$ and $c_0 = 2\mpi^2/(\fpi e)^2$. 
The energy momentum tensor of the model can be obtained via the Hilbert prescription, which for our sign convention is given by
\begin{equation}
    T^{\mu\nu}=-\frac{2}{\sqrt{-g}}\frac{\delta S}{\delta g_{\mu\nu}}=-\frac{2}{\sqrt{-g}}\frac{\partial(\lag \sqrt{-g})}{\partial g_{\mu\nu}}
\end{equation}
so that we have
\begin{align}
    \notag T_{\mu\nu}=&\frac{1}{2}\Tr{-2L_\mu L_\nu+\eta_{\mu\nu}L_\rho L^\rho}-\frac{1}{4}\Tr{-4[L_\mu,L_\rho] [L_\nu,L^\rho]+\eta_{\mu\nu}[L_\rho,L_\sigma]^2} \\
    &+8\pi^4c_6 B_\mu B_\nu -4\pi^4c_6 B_\rho B^\rho\eta_{\mu\nu}  +c_0(1-\sigma)\eta_{\mu\nu}.
    \label{classical}
\end{align}
\comment{
The Hilbert prescription with our conventions yields
where, for the sextic term in a gravitational field
\begin{equation}
    S=-\int d^4x\sqrt{-g}\lambda^2\pi^4g_{\mu\nu}\frac{B^\mu B^\nu}{-g}
\end{equation}
}
Note that the definition of $B^\mu$ involves a Levi-Civita tensor, hence it transforms as a tensor density of rank one and each must include a factor of $|g|^{-\tfrac{1}{2}}$ when coupled to a nontrivial gravitational field \cite{Adam:2015BPSlimit}. 
\comment{
Then, we have
\begin{equation}
    -\frac{2}{\sqrt{-g}}\frac{\delta(\lag \sqrt{-g})}{\delta g_{\mu\nu}}=\lambda^2\pi^4\frac{2}{\sqrt{-g}}\qty[\frac{B^\mu B^\nu }{\sqrt{-g}}+g_{\alpha\beta}B^\alpha B^\beta\frac{\delta |g|^{-\tfrac{1}{2}}}{\delta g_{\mu\nu}}]=2\lambda^2\pi^4\qty[\frac{B^\mu B^\nu }{|g|}-\frac{1}{2|g|}g_{\alpha\beta}B^\alpha B^\beta g^{\mu\nu}]
\end{equation}
and $T^{00}$ has overall positive sign. To compute $T_{00}$ with Hilbert prescription one should take into account that $B_\mu =g_{\mu\nu}B^\nu$ also.
\\}
\comment{{\color{red}
Let me proceed differently. I can eliminate the Levi-Civita tensor as 
\begin{eqnarray}
L\sim - B^2 = \frac{3}{2(24\pi^2)^2}{\rm Tr}[[L_\alpha, L_\beta] L_\gamma]{\rm Tr}[[L^\alpha, L^\beta] L^\gamma] 
\end{eqnarray}
From this I get 
\begin{equation}
T_{\mu\nu} =  \frac{9}{(24\pi^2)^2}{\rm Tr}[[L_\mu ,L_\beta] L_\gamma]{\rm Tr}[[L_\nu, L^\beta] L^\gamma] -\eta_{\mu\nu}L 
\end{equation} 
Therefore
\begin{eqnarray}
T_{00} &=& -L  \\
 T_{ij}&=& \frac{9}{(24\pi^2)^2}{\rm Tr}[[L_i ,L_k] L_l]{\rm Tr}[[L_j, L^k] L^l] +\delta_{ij}L \\
 &=& -\delta_{ij}  \frac{3}{(24\pi^2)^2}{\rm Tr}[[L_m ,L_k] L_l]{\rm Tr}[[L^m, L^k] L^l] +\delta_{ij} L = -\delta_{ij}L
\end{eqnarray}  
where I used that because of cyclicity, ${\rm Tr}[[L_i ,L_k] L_l]$ must be proportional to $\epsilon_{ikl}$. This agrees with (6)! 
}
}

We shall be mainly interested in the $\mu\nu=i,j$ (spatial) components. The relevant traces we need to compute are:
\begin{align}
    \Tr{-L_i L_j}&=\Tr{\partial_i U^\dagger \partial_jU}=2\partial_i\phi^\alpha \partial_j\phi^\alpha, \\[2mm]
    \Tr{L_i L_jL_kL_l}&=\Tr{\partial_i U^\dagger\partial_jU\partial_kU^\dagger\partial_lU} \notag\\
    &=2\partial_i\phi^\alpha\partial_j\phi^\beta\partial_k\phi^\gamma\partial_l\phi^\delta(2\delta_{\alpha\beta}\delta_{\gamma\delta}+\delta_{\alpha\gamma}\delta_{\beta\delta}-\delta_{\alpha\delta}\delta_{\beta\gamma}-\epsilon_{\alpha\beta\gamma\delta})
\end{align}
and contractions of these. We have used the properties of the $\tau_\alpha$ (see e.g. appendix D in \cite{Espinosa:1989qn}).
Classical solutions are obtained as minimizers of the static energy functional \eqref{Energy_integral}, via the (accelerated) gradient flow algorithm (see \cref{sec:numerics}).

\subsection{Quantization}
Nucleons and nuclei are described within the Skyrme model as classical solitonic configurations through the identification of the Skyrmion topological charge and the baryon number of nuclear states. However, other quantum numbers such as the spin and isospin of quantum nuclear states are not described at the classical level. Hence, a quantization of the Skyrmion field is needed in order to take into account the relevant quantum numbers. This is done in the semi-classical approach by promoting the zero modes of the soliton to dynamical degrees of freedom.

To do so, we introduce the rotational and iso-rotational degrees of freedom through a pair of time-dependent $SU(2)$ transformations of the classical (static) solitonic solution, 
representing the iso-rotation and the spatial rotation zero modes, respectively, as well as a time dependent vector $\vec{X}(t)$ representing the translational zero modes:
\begin{equation}
    U(t,\bm{x})= A(t) U_0(R_B(t)(\bm{x}-\bm{X}(t)))A^\dagger (t)
    \label{transformiso}
\end{equation}
where $R^{ij}_B=\tfrac{1}{2}\Tr{\tau^i B\tau^jB^\dagger }\in SO(3)$ is the corresponding rotation matrix in space.  $A(t),B(t)\in SU(2)$ 
and ${\bm X}(t)$ together form the \emph{collective coordinates} of the soliton.\footnote{For  spherically symmetric solutions,  rotation in  coordinate space can be undone by that in  isospin space, and one can devise a simpler quantization procedure without introducing the matrix $B(t)$  \cite{Adkins:1983ya}. The present treatment is more general and can be used also for  non-spherical solutions that we shall be mainly interested in. }  The semi-classical quantization of the Skyrmion then consists of substituting \eqref{transformiso} into the Skyrme Lagrangian \eqref{Lagrangian}, which yields the Lagrangian of an effective dynamical system in terms of the collective coordinates $\{A(t),B(t),\bm{X}(t)\}$, and quantizing such a system via standard canonical methods. Performing this substitution yields
\begin{equation}
    L_{\rm col}=\int d^{3} x \mathcal{L}_{\rm SK}=-\mathcal{M}+\frac{1}{2} \mathcal{M} \dot{X}_{i} \dot{X}_{i}+\frac{1}{2} a_{i} U_{i j} a_{j}-a_{i} W_{i j} b_{j}+\frac{1}{2} b_{i} V_{i j} b_{j},
    \label{collectivelag}
\end{equation}
where 
$$
a_{j}=-i \operatorname{Tr} \tau_{j} A^{-1} \dot{A}, \quad b_{j}=i \operatorname{Tr} \tau_{j} \dot{B} B^{-1},
$$
are the angular velocities in isospace and physical space, respectively. We have also introduced the corresponding inertia tensors 
\begin{align}
U_{i j} &=\frac{1}{24\pi^2}\int d^{3} x \qty[-\Tr{T_{i} T_{j}}-\Tr{\left[T_{i},L_{k}\right]\left[T_{j},L_{k}\right]} + \frac{c_6}{16}\Tr{T_i[L_a,L_b]}\Tr{T_j[L_a,L_b]}],  \label{eq:inertiatens}\\[2mm]
V_{i j} &=\frac{1}{24\pi^2}\int d^{3} x\, \epsilon_{i l m} \epsilon_{j n p} x_{l} x_{n} \qty[-\Tr{L_p L_m} - \Tr{\left[L_p,L_{k}\right]\left[L_m,L_{k}\right]} + \frac{c_6}{16}\Tr{L_p[L_a,L_b]}\Tr{L_m[L_a,L_b]}], \label{eq:spintens} \\[2mm]
W_{i j} &=\frac{1}{24\pi^2}\int d^{3} x \, \epsilon_{j l m} x_{l} \qty[\Tr{T_{i}L_m}+\Tr{\left[T_{i},L_{k}\right]\left[L_m,L_{k}\right]} - \frac{c_6}{16}\Tr{T_i[L_a,L_b]}\Tr{L_m[L_a,L_b]}] ,\label{eq:mixedtens}
\end{align}
given in terms of  the $\mathfrak{s u}(2)$-valued  currents
\begin{equation}
    L_k=U_0^\dagger \partial_k U_0,\qquad T_{i}=\frac{i}{2}U_{0}^{\dagger}\left[\tau_{i},U_{0}\right].
\end{equation}
 For notational simplicity, here and below we do not distinguish upper ($x^{i},\tau^a$) from lower ($x_{i},\tau_a$) indices for three-dimensional vectors and tensors when we deal with purely three-dimensional expressions. 

Once we have the Lagrangian \eqref{collectivelag} (and thus the corresponding Hamiltonian), we may perform the quantization of the system by finding an irreducible representation of the algebra of observables associated to the quantum degrees of freedom defined on the Hilbert space of states, $\mathcal{H}$. 
We start by noting that the term quadratic in the time derivatives of the translational coordinates $X_i$ in \eqref{collectivelag} is the standard kinetic term of a non-relativistic free particle. The associated observables will be the translational coordinates $X_i$ and momenta $P_i=\mathcal{M} \dot X_i$, and the corresponding quantum operators satisfy the standard canonical commutation relations,
\begin{equation}
    [\hat{X}_i,\hat{P}_i]=i\delta_{ij}.
\end{equation}
Hence, the corresponding states associated to the translational zero modes will be those of a quantum, non-relativistic free particle. Let $\mathcal{H}_{\rm f.p.}$ be the Hilbert space of such states. A complete basis for $\mathcal{H}_{\rm f.p.}$ is given by the set of momentum eigenstates, $\ket{\bm{p}}$.

On the other hand, a transformation of the form \eqref{transformiso} with $\bm{X}=0$ corresponds to an element $g=(A,B)\in SU(2)\times SU(2)\equiv G$ of the most general symmetry group of the Skyrme Lagrangian. Therefore, the quantum mechanical spin and isospin states of a Skyrmion belong to the Hilbert space $\mathcal{H}_G=L^2(G,\mu)$ of square integrable functions on $G$ \footnote{Actually, on some cover of $G$ \cite{Krusch:2006tg}.}
with the inner product
\begin{equation}
    \ev{\psi,\phi}=\int_G d\mu(g)\psi^*(g)\phi(g),
    \label{innerprod}
    \end{equation}
where $d\mu(g)$ is the Haar measure on $G$.
The observables associated to these states correspond to the rotational and iso-rotational collective coordinates $A,B$ and their canonically conjugate momenta, the body-fixed isospin and spin angular momentum operators $K_{j}$ and $L_{j}$, obtained in terms of the angular velocities $a_{j}$ and $b_{j}$ via the relations,
\begin{equation}
\begin{aligned}
K_{i} &=U_{i j} a_{j}-W_{i j} b_{j} \\
L_{i} &=-W_{i j}^{T} a_{j}+V_{i j} b_{j}
\label{momentum-velocities}
\end{aligned} 
\end{equation}
where $T$ denotes transpose. These operators are related to the usual space-
fixed isospin and spin angular momentum operators $I_{j}$ and $J_{j}$ via
\begin{equation}
I_{i}=-R_{i j}(A) K_{j}, \quad J_{i}=-R_{i j}\left(B\right)^{T} L_{j}
\label{body-fixed-relations}
\end{equation}
implying $\bm{I}^{2}=\bm{K}^{2}$ and $\bm{J}^{2}=\bm{L}^{2}$. The set of operators, $\bm{I}, \bm{J}, \bm{K}$ and $\bm{L},$ form an irreducible representation of the Lie algebra of $\mathcal{O}_{I, K} \otimes \mathcal{O}_{L_{n} J},$ the symmetry group of two rigid rotators, and obey
the commutation relations
\begin{equation}
{\left[I_{i}, I_{j}\right]=i \epsilon_{i j k} I_{k},}\quad  {\left[K_{i}, K_{j}\right]=i \epsilon_{i j k} K_{k}} ,\quad
{\left[J_{i}, J_{j}\right]=i \epsilon_{i j k} J_{k 1}} ,\quad {\left[L_{i}, L_{j}\right]=i \epsilon_{i, j k} L_{k}}.
\end{equation}
From these, we may produce a complete set of commuting observables to define an eigenstate basis of $\mathcal{H}_G$. We construct such a basis with states of the form
\begin{equation}
    \ket{G}=\ket{i,i_3,k_3}\otimes\ket{j,j_3,l_3}\equiv\ket{i,i_3,k_3;j,j_3,l_3}
    \label{basisG}
\end{equation}
where $j$ and $j_3$ correspond to the eigenvalues of the corresponding total angular momentum and the third component of angular momentum operators. 
Thus the total Hilbert space $\mathcal{H}=\mathcal{H}_{\rm{f.p.}}\otimes \mathcal{H}_G$ is spanned by the states $\ket{i, i_{3}, k_{3} ; j, j_{3}, l_{3} ; \bm{p}},$ where $-i \leq i_{3}, k_{3} \leq i$ and $-j \leq j_{3}, l_{3} \leq j .$ In particular, the subspace of fixed $i, i_{3}, j, j_{3}$ and momentum $\bm{p},$ labelled by the states $\ket{k_{3}, l_{3}},$ is $(2 i+1)(2 j+1)-$dimensional.

\section{The D term}

In this section, we introduce the gravitational form factors  and outline our strategy to compute the $D(t)$ form factor for the $B=1$ (nucleons), $B=2$ (deuteron) and $B=3$ (helium-3 and tritium)  solutions. These three examples are representative of different situations one  encounters in the computations of $D(t)$ in the Skyrme model. Despite the differences, however, under certain assumptions we shall arrive at the same formula in all cases. This motivates us to extrapolate our discussion to solutions with arbitrary values of $B$. The actual computation of the form factor is carried out in  the next section. 


\subsection{$B=1$: the nucleons}

We start with the nucleons (proton and neutron) with $B=1$. 
The off-forward  (i.e., nonzero momentum transfer) nucleon matrix element of the QCD energy momentum tensor can be parameterized as 
\begin{equation}
    \langle p'|T_{\mu\nu}|p\rangle=\Bar{u}(p')\Big[\gamma_{(\mu }P_{\nu)} A(t)+\frac{iP_{(\mu}\sigma_{\nu)\rho}q^\rho}{2M_N}B(t)+\frac{q_\mu q_\nu -\eta_{\mu\nu}q^2}{4M_N}D(t)
    \Big]u(p) \label{nucleon}
\end{equation}
where $P_\mu=(p_\mu+p'_\mu)/2$ and  $q=p'-p$ is the momentum transfer with $t=q^2$.  $A_{(\mu}B_{\nu)}=\frac{A_\mu B_\nu+A_\nu B_\mu}{2}$ denotes symmetrization.  $u(p)$ is the nucleon spinor normalized as $\Bar{u}(p)u(p)=2M_N$ with $M_N$ being the mass of the nucleon. The same formula applies to all the spin-1/2 nuclei with trivial changes. The gravitational form factors $A,B,D$ are scalar, renormalization-group invariant functions of $t$. In the zero momentum transfer limit, i.e. $t=0$, the values of $A(t=0)$ and $B(t=0)$ are constrained to $1$ and $0$ due to momentum and angular momentum conservation, respectively. However, the so-called D-term  $D=D(t=0)$ is not constrained by any symmetry,  and this is our main object of interest.

The $D(t)$ form factor can be isolated by  working in the Breit frame (${\bm{p}}'=-{\bm p}={\bm q}/2$, $t=-\bm{q}^2$) and taking the spatial $i,j=1,2,3$ components 
\begin{equation}
    \langle p'|T_{ij}|p\rangle=2P^0\frac{D(t)}{4M_N}(q_iq_j-\bm{q}^2\delta_{ij}) \approx \frac{D(t)}{2}(q_iq_j- \bm{q}^2\delta_{ij}),
    \label{Breit}
\end{equation}
 where we used the nonrelativistic approximation in the last expression. Classically, in the Skyrme model, $D(t)$ can be obtained by simply Fourier-transforming the classical energy momentum tensor $T^{cl}_{ij}$ (\ref{classical}). For spherically symmetric configurations such as the $B=1$ solution,  the  energy momentum tensor can be written in terms of the `shear' and `pressure' distributions
\begin{equation}
    T^{cl}_{ij}({\bm x})=\qty(\frac{x_ix_j}{{x^2}}-\frac{1}{3}\delta_{ij})s(x)+p(x)\delta_{ij}.
    \label{symcon}
\end{equation}
(Below we shall write $|\bm{x}|=x$ and $|\bm{q}|=q$ for simplicity.) 
It then follows that 
\begin{eqnarray}
  D(t)&=&-6M_N\int d^3x \left(x^ix^j-\frac{1}{3}\delta^{ij} x^2\right)\frac{j_2(qx)}{(qx)^2}T^{cl}_{ij}({\bm x}) . \label{main}
\end{eqnarray}
In particular, 
\begin{eqnarray}
    D(t=0)&=&-\frac{2M_N}{5}\int d^3 x\left(x_ix_j-\frac{1}{3} x^2\delta_{ij}\right)  T^{cl}_{ij}({\bm x})\nonumber \\
    &=& -\frac{4M_N}{15} \int d^3x\, x^2s(x). \label{ss}
\end{eqnarray}
We see that the D-term is related to the distribution of shear forces inside the nucleon, parametrized in the spherically symmetric case by the function $s(x)$. 
It is interesting to note that in the so-called BPS  Skyrme model, a solution that saturates the Bogomolny  bound exists \cite{Adam:2010ds}. For this particular solution, the energy momentum tensor is that of a perfect liquid \cite{Adam:2015rna}, that is, the shear force vanishes. Therefore, the D-term is exactly zero in this  limit. 
From physical point of view, the passing of a gravitational wave causes a volume preserving deformation of the nucleon, which is precisely a symmetry in the BPS model \cite{Adam:2010ds}. Therefore, in this limit, nucleons are transparent to gravitons.

Returning to  general situations with $s(x)\neq 0$, we now consider the effect of quantization (\ref{transformiso}) and write the matrix element between momentum eigenstates as
 \begin{equation}
     \langle \bm{q}/2|T_{ij}[ U(R(B)(\bm{x}-\bm{X}))] |-\bm{q}/2\rangle=e^{-i\bm{q}\cdot \bm{x}}R_{ia}^T(B)R_{jb}^T(B)\int d^3x'\exp{i\bm{q}\cdot R^T(B)\bm{x}'}T_{ab}(\bm{x}'),
     \label{Breit_matelements}
 \end{equation}
where 
\begin{equation}
    T_{ab}(\bm{x})=T^{cl}_{ab}[U_0(\bm{x})]+\order{I^2,J^2}. \label{quant}
\end{equation}
Both $R^T$ and $T_{ab}$ in (\ref{Breit_matelements}) are operators which act on spin and isospin eigenstates (\ref{basisG}). In general, the calculation of their matrix elements is a complicated task. However, since our main objective is to evaluate the $D(t)$-form factor for a wide variety of nuclei with different spin/isospin quantum numbers rather than focusing on a particular nucleus, we neglect the ${\cal O}(I^2,J^2)$ terms in (\ref{Breit_matelements}). This may be partially justified by the large-$N_c$ approximation. 

 Even after this approximation, (\ref{Breit_matelements}) still contains the spatial rotation matrix $R^T$  which acts on external spin states. To deal with it, we expand the exponential  on the right hand side in partial waves $l=0,1,2,\cdots$ 
 \begin{eqnarray}
    \exp{i\bm{q}\cdot R^T(B)\bm{x}}&=&j_0(qx)+iq_cR_{ck}^T(B)x_k\frac{3j_1(qx)}{qx} \\
   &&-\frac{1}{2}\left(q_cq_d-\frac{1}{3}\delta_{cd} q^2\right) R_{ck}^T(B)R_{dl}^T(B)\left(x_kx_l-\frac{1}{3}\delta_{kl} x^2\right)\frac{15j_2(qx)}{(qx)^2}+\cdots. \notag
     \label{part}
 \end{eqnarray}
In order to extract the $D(t)$-form factor, it suffices to focus on the $l=2$ tensor term and read off the coefficient of  $q_iq_j$. 
One then needs to evaluate the integral 
\begin{eqnarray}
T_{abkl}=\int d^3\bm{x} \left(x_kx_l-\frac{1}{3}\delta_{kl} x^2\right)\frac{15j_2(qx)}{(qx)^2}T^{cl}_{ab}(\bm{x}) \label{4t}
\end{eqnarray}
For a spherically symmetric solution, the tensorial structure is completely fixed by symmetry 
\begin{equation}
T_{abkl}= \frac{1}{10}\left(\delta_{ak}\delta_{bl}+\delta_{il}\delta_{jk}-\frac{2}{3}\delta_{ij}\delta_{kl}\right)T_{cdcd}.
\label{tt}
\end{equation}
Note that the trace part $T^{cl}_{ab}\sim \delta_{ab}$ does not contribute to the integral  (\ref{4t}).  
Substituting (\ref{tt}) into (\ref{Breit_matelements}), we see that the $R^T$ matrices disappear due to the orthogonality relation $RR^T=1$ and we recover the formula (\ref{main}). 

One might wonder that, since we have neglected the ${\cal O}(J^2,I^2)$ terms in (\ref{quant}), what the effects of quantization are in the present calculation.  The point is that if the classical solution is not spherically symmetric, the tensor $T_{abkl}$ does not have the canonical form (\ref{tt}) in general. The introduction of the $R$-matrix then becomes crucial to restore the symmetry. We shall see examples of this below. 

\subsection{$B=2$: the deuteron}

Next we consider the $B=2$ sector, the `deuteron', with spin $J=1$.  For spin-1 nuclei, there are in general six independent  gravitational form factors \cite{Holstein:2006ud} related to the fact that the nuclear wavefunction is not spherically symmetric. In particular, it is well-known that the deuteron wavefunction has quadrupole deformation.  In terms of the energy momentum tensor matrix element,  this is most clearly shown by the following multipole expansion  in the Breit frame   in the  nonrelativistic limit \cite{Polyakov:2019lbq}
\begin{align}
\langle p'\sigma'|T_{ij}|p \sigma\rangle = \frac{1}{2}(q_i q_j -\delta_{ij}q^2) {\cal D}_1(t)\epsilon^*_{\sigma'}\cdot \epsilon_\sigma + \left(q_j q_k Q_{ik} + q_i q_k Q_{jk} -q^2Q_{ij} -\delta_{ij} q_k q_l Q_{kl}\right)_{\sigma'\sigma}{\cal D}_2(t) \notag \\ + \frac{1}{2M_D^2}(q_i q_j-\delta_{ij}q^2) q_k q_l Q_{kl,\sigma'\sigma} {\cal D}_3(t)+\cdots
\end{align}
where $M_D$ is the deuteron mass. $\epsilon_\sigma$  are the polarization vectors of the deuteron with spin $\sigma= \pm 1,0$ measured  along the $x^3$-direction (not helicity). $Q_{ij,\sigma'\sigma}\equiv \langle \sigma'|\hat{Q}_{ij}|\sigma\rangle$ is the matrix element of the quadrupole operator
\begin{equation}
\hat{Q}_{ij}=\frac{1}{2}\left(J_iJ_j+J_jJ_i -\frac{4}{3}\delta_{ij}\right),
\end{equation}
where $J_i$ is the spin-1 operator with matrix elements $\langle \sigma'|J_i|\sigma\rangle = -i\epsilon_{ijk} \epsilon_{\sigma' j}^*\epsilon_{\sigma k}$. 
While the quadrupole part is of interest in its own right (see recent extractions for the $\rho$-meson \cite{Freese:2019bhb,Pefkou:2021fni}), in this work we focus on the monopole part ${\cal D}_1$. The $Q$-dependent terms can be eliminated by averaging over the three spin states thanks to the identity 
\begin{equation}
\sum_{\sigma}^{\pm, 0} Q_{ij,\sigma\sigma}=0. 
\end{equation}
Restricting ourselves to this simpler situation, we write 
\begin{equation}
\frac{1}{3}\sum_{\sigma}\langle p'\sigma|T_{ij}|p \sigma\rangle = \frac{1}{2}(q_i q_j -\delta_{ij}q^2)D(t),
\label{deut} 
\end{equation}
where we renamed $D(t)={\cal D}_1(t)$ to emphasize the correspondence with (\ref{Breit}).

Turning now to the Skryme model, we recall that the classical $B=2$ solution  has toroidal symmetry 
\cite{Braaten:1988cc} and is mainly characterized by a c-number quadrupole tensor $Q_{ij}$ \cite{Polyakov:2019lbq}
\begin{eqnarray}
T^{cl}_{ij}\approx Y_2^{ij} s(x) + p(x)\delta_{ij} &+& 2s'(x)\left(Q_{ik}Y_2^{kj}+Q_{jk}Y_2^{ki}-\delta_{ij}Q_{ab}Y_2^{ab}\right) + p'(x)Q_{ij}
\nonumber \\
&& -\frac{1}{M_D^2}Q^{kl}\partial_k \partial_l \left(p''(x)\delta^{ij}+s''(x)Y_2^{ij}\right),
\label{ij}
\end{eqnarray}
where we abbreviated $Y_2^{ij}=\frac{x_ix_j}{x^2}-\frac{\delta_{ij}}{3}$. In principle, $Q_{ij}$ can be  numerically extracted from a given Skyrmion configuration. But in parallel with the spin-averaging procedure above, we eliminate it by forming the moment (\ref{main}) and using the integrals 
\begin{eqnarray}
&&\int d^2\Omega Y_2^{ij}=0, \qquad \int d^2\Omega Y_2^{ij} \left(Q_{ik}Y_2^{kj}+Q_{jk}Y_2^{ki}-\delta_{ij}Q_{ab}Y_2^{ab}\right) =0, \nonumber \\
&&  Q^{kl} \int d^2\Omega Y_2^{ij} \partial_k \partial_l\left(s''(x)Y_2^{ij}\right)=0,
\end{eqnarray}
where we used the traceless property $Q_{ii}=0$. We thus see that the $D(t)$ form factor as defined in (\ref{deut}) can be calculated via the same formula (\ref{main}) with the trivial replacement $M_N\to M_D$ even if the classical solution is not spherical.

\subsection{$B=3$ and beyond}

Next we turn to the $B=3$ sector relevant to the helium-3 nucleus $(^3$He) and the tritium $(^3$H). Since they have spin $1/2$, GFFs are parameterized  by the same formula (\ref{nucleon}) as in the nucleon case. However, in the Skyrme model, the $B=3$ classical solutions are not spherically symmetric, but rather have the shape of a tetrahedron \cite{Carson:1990yv} (see \cref{fig:1_8} below). It is then not obvious how one can recover  the same set of form factors in the present semiclassical approach. 

In order to answer this question, we need some elements of group theory.  The tetrahedral group $T_d$ \cite{Tinker} consists of 24 discrete transformations such as a  120$^{\rm o}$ rotation around one of the four vertices of a tetrahedron. 
It turns out  that this symmetry  imposes strong constraints on various moments of  classical configurations \cite{Carson:1991fu}. For example, consider  the following integral
\begin{equation}
A_{ab}= \int d^3x T^{cl}_{ab}(\bm{x}) j_0(qx),
\end{equation}
which appears in the $l=0$ partial wave in (\ref{part}). Let $g$ be an element of the tetrahedral group. Changing variables as ${\bm x}\to g{\bm x}$, we find 
\begin{equation}
A_{ab}= g^T_{ai}g^T_{bj}\int d^3x' T'^{cl}_{ij}(\bm{x}') j_0(qx') = g^T_{ai}g^T_{bj}A_{ij}
\end{equation}
where we used $T'^{cl}_{ij}=T^{cl}_{ij}$ due to symmetry. This means that $A_{ab}\propto \delta_{ab}$. Mathematically, the integral transforms as the  singlet $(A_1)$  representation contained in the product of two vector representations ($T_1$) of the tetrahedral group  
\begin{equation}
T_1\times T_1 = A_1+E+T_1+T_2,
\end{equation}
along with the two-dimensional ($E$) and axial vector ($T_2$) representations \cite{Tinker}. 
 
 
Consider, then, the transformation properties of the tensor $T_{abkl}$ defined in (\ref{4t}) under the tetrahedral group. Since the trace  part of the energy momentum tensor $T_{ab}^{cl}$ does not play a role,  $T_{abkl}$ can be viewed as the product of two symmetric and traceless tensors formed by $T_1$ vectors
\begin{equation} 
(T_1\times T_1)\times (T_1\times T_1),
\end{equation}
where $(T_1\times T_1)=E+T_1$ denotes the traceless part. 
We are interested in the components that transform as the irreducible, singlet representation $A_1$. There are two such structures\footnote{ One coming from $E\times E=A_1+A_2+E$, the other from $T_1\times T_1=A_1+E+T_1+T_2$. The cross term $E\times T_1=T_1+T_2$ does not contribute to the trivial representation.} which can be rearranged in the form \cite{Carson:1991fu} 
\begin{align}
T_{abkl} = \frac{1}{10}\left(\delta_{ak}\delta_{bl}+\delta_{al}\delta_{bk}-\frac{2}{3}\delta_{ab}\delta_{kl}\right)T_{cdcd} +C_{abkl}   .\label{C}
\end{align}
The first structure is the same as before (\ref{tt}). The second tensor $C$ is symmetric and traceless in any pair of indices $(a,b,k,l)$ (see \cite{Carson:1991fu} for the details). Plugging (\ref{C}) into (\ref{Breit_matelements}), we obtain  
\begin{eqnarray}
    \left.\bra{\bm{q}/2}T_{ij}(-R{\bm X})\ket{-\bm{q}/2}\right|_{l=2} 
    &=& -\left(q_iq_j-\frac{\delta_{ij}}{3}q^2\right)\frac{T_{cdcd}}{10}  \notag \\
    && -\frac{1}{2}\left(q_cq_d-\frac{1}{3}\delta_{cd} q^2\right) R_{ia}^T(B)R_{jb}^T(B)R_{ck}^T(B)R_{dl}^T(B)C_{ab,kl} . \label{C1}
\end{eqnarray}
The operator in  the second term is totally symmetric and traceless in the four indices $i,j,c,d$. It is thus a spin-4 operator whose matrix element between spin-1/2 states vanishes.  The first term then leads to the same formula (\ref{main}) for the $D$-form factor. We now appreciate the effect of the rotation matrix $R$ in the present calculation. Had we neglected $R$'s, namely, $R_{ia}^T\to \delta_{ia}$ etc. in (\ref{C1}), the second term would have contributed extra terms quadratic in $q_i$, in contradiction with the unique tensor structure (\ref{Breit}) for a spin-1/2 nucleus. By introducing $R$'s, we  have minimally included quantum effects in order to restore  the original spherical symmetry of the problem.  \\

The above argument can be broadly generalized. Each Skyrmion solution possesses a symmetry group \cite{BjarkeGudnason:2018bju}. For $B\ge 3$, the group consists of discrete symmetry transformations and this puts strong constraints on the possible tensor structures of the integral (\ref{4t}). For example, the $B=4$ solution (the helium-4 nucleus, or the `alpha' particle)  has cubic symmetry \cite{Braaten:1989rg}. The associated octahedral group $O_h$ has irreducible representations which correspond to those of the tetrahedral group $A_1\to A_{1g}$, $T_1\to T_{1u}$, etc. We can then immediately  conclude that the $D$-form factor is again given by (\ref{main}).\footnote{For a spinless nucleus such as the helium-4, there are only two GFFs which we parameterize  as  
\begin{equation}
\langle p'|T_{\mu\nu}|p\rangle = 2A(t)P_{\mu}P_{\nu} + \frac{D(t)}{2}(q_\mu q_\nu - \eta_{\mu\nu}q^2). 
\end{equation}}
A similar argument can be repeated for larger-$B$ solutions. (Spin-averaging is understood for spin $\ge 1$ nuclei.)  We thus use (\ref{main}), with $M_N$ replaced by nuclear masses, as a working definition for all nuclei in the Skyrme model.    

\section{Angular momentum form factor $J(t)$ }

Although our main interest in this paper is the $D(t)$-form factor, our approach can be straightforwardly generalized to the other gravitational form factors. As an example,  let us consider the form factor 
\begin{equation}
J(t)=\frac{1}{2}(A(t)+B(t)),
\end{equation}
for spin-$1/2$ nuclei  where $A(t),B(t)$ are  defined in (\ref{nucleon}). Physically, $J(t)$ represents the form factor associated with the total angular momentum of the system. The forward value $J(0)=1/2$ is constrained by angular momentum conservation. In the Skyrme model, $J(t)$ has been computed for the $B=1$ solutions \cite{Cebulla:2007ei,Kim:2020lrs}. Here,  for the first time, we compute it for the $B=3$ (helium-3, tritium) solution. 

In the Breit frame, the $J$-form factor appears in the following components of the energy momentum tensor matrix element \eqref{nucleon}

\begin{align} & \frac{\bra{p^{\prime}, s^{\prime}}T_{00}(0)\ket{ p, s}}{2P^0}= M \left[A(t)-\frac{t}{4 M^2}\left[A(t)-2 J(t)+D(t)\right]\right] \delta_{s s^{\prime}} \\ 
              & \frac{\bra{p^{\prime}, s^{\prime}}T_{0 i}(0)\ket{ p, s}}{2P^0}=-  J(t) \, i\epsilon_{ijk}\frac{\tau^j_{s's}}{2} q^k. \label{mat}
\end{align} 
Note that the mixed (timelike-spacelike) components vanish in the classical limit which corresponds to a static configuration. Once the quantization of the Skyrmion is taken into account, we find a nonzero, operator-valued result
\begin{equation}
    T_{0i}=I_{ij}a^j-J_{ij}b^j, \label{four}
\end{equation}
where
\begin{align}
    I_{ij}&=-\frac{1}{24\pi^2}\qty[ \Tr{L_{i} T_{j}}+\Tr{\left[L_{k}, L_{i}\right]\left[L_{k}, T_{j}\right]}] ,\label{eq:icurr}\\[2mm]
    J_{ij}&=\frac{1}{24\pi^2}\epsilon_{jkl}x^k\qty[ \Tr{L_{i} L_{l}}+\Tr{\left[L_{a}, L_{i}\right]\left[L_{a}, L_{l}\right]}].\label{eq:jcurr}
\end{align}
Both of the above tensor densities transform as the product of a vector and an axial vector, i.e., they  belong to the $T_2\times T_1$ representation which does not contain an irreducible trivial component $A_1$. 
Therefore, their contribution must vanish at first order ($l=0$) in the partial wave expansion (\ref{part}). At the next order $l=1$, we find
\begin{align}
    \left.\langle \bm{q}/2|T_{0i}(-R\bm{X}) |-\bm{q}/2\rangle\right|_{l=1}=&\,R_{ij}^T(B)\int d^3x\exp{i\bm{q}\cdot R^T(B)\bm{x}}T_{0j}(\bm{x})
   \notag \\
    =&\,iR_{ij}^T(B)R_{kl}^T(B)q^k\int d^3x \frac{3j_1(qx)}{qx}\qty[I_{jm}a^m-J_{jm}b^m]x^l . \label{Breit_matelements_0j}
\end{align}
Defining the currents 
\begin{equation}
    I_{ijk}(q)=\int d^3x\frac{3j_1(qx)}{qx}I_{ij}x^k,\qquad J_{ijk}(q)=\int d^3x\frac{3j_1(qx)}{qx}J_{ij}x^k,
\end{equation}
we can identify their irreducible component that transforms as $A_1\in T_2\times T_1\times T_1$, which is totally antisymmetric in the three indices \cite{Carson:1991fu}:
\begin{equation}
    I_{ijk}= \epsilon_{ijk}\mathcal{I}(q),\qquad J_{ijk}= \epsilon_{ijk}\mathcal{J}(q), \label{only}
    \end{equation}
where
    \begin{equation}
\mathcal{I}(q)=\frac{1}{3!}\epsilon^{abc}I_{abc}(q), \qquad \mathcal{J}(q)=\frac{1}{3!}\epsilon^{abc}J_{abc}(q).
    \label{IJdefs}
\end{equation} 
Substituting \eqref{only} into \eqref{Breit_matelements_0j}, we get 
\begin{align}
    \langle \bm{q}/2|T_{0i}(-R\bm{X}) |-\bm{q}/2\rangle=&\,iR_{ij}^T(B)R_{kl}^T(B)q^k\epsilon_{jml}\qty[\mathcal{I}(q)a^m-\mathcal{J}(q)b^m] \notag\\[2mm]
    =&\,i\epsilon_{ijk}q^kR_{jm}^T(B)\qty[\mathcal{I}(q)a^m-\mathcal{J}(q)b^m].
\end{align}
At this point, we need to invert the relations in \eqref{momentum-velocities} between the spin and isospin velocities in terms of the associated angular momentum operators. 
This is a complicated task in general, but for the specific $B=3$ solution at hand,  
the tensors defined in 
\cref{eq:inertiatens,eq:spintens,eq:mixedtens}  
are proportional to the identity, i.e., $U_{ij}=u\delta_{ij},V_{ij}=v\delta_{ij},W_{ij}=w\delta_{ij} $. In that case, we have \cite{Carson:1991fu}:
\begin{equation}
    a^m=\frac{v K^m+w L^m}{uv-w^2},\qquad b^m=\frac{w K^m+u L^m}{uv-w^2},
\end{equation} 
and we may write
\begin{equation}
    R_{lm}^T(B)\qty[\mathcal{I}(q)a^m-\mathcal{J}(q)b^m]=\frac{1}{uv-w^2}\qty{\qty[\mathcal{I}(q)(w-v)+\mathcal{J}(q)(w-u)]R^T_{lm}(B)L^m+\qty[v\mathcal{I}(q)-w\mathcal{J}(q)] R_{lm}(B)M^m},
\end{equation}
where we have defined the operator $\bm{M}=\bm{K}+\bm{L}$. Also, taking into account \eqref{body-fixed-relations} and the fact that 
the $B=3$ ground states are $\bm{M}=0$ singlets  \cite{Carson:1990yv}, we find our final result 
\begin{equation}
    J(t)=\frac{\mathcal{I}(q)(w-v)+\mathcal{J}(q)(w-u)}{uv-w^2}.
\end{equation}
Comparing the definitions in \cref{IJdefs,eq:icurr,eq:jcurr} with \cref{eq:mixedtens,eq:spintens}, we get the following relations
\begin{equation}
    \lim_{q\rightarrow 0}\mathcal{I}(q)=-\frac{1}{6}W_{ii}=-\frac{1}{2}w,\quad \lim_{q\rightarrow 0}\mathcal{J}(q)=-\frac{1}{6}V_{ii}=-\frac{1}{2}v,
\end{equation}
and thus $J(t=0)=\frac{1}{2}$ as required by Lorentz invariance. 

\section{Numerical results}
\label{sec:numerics}
\begin{figure}
     \centering
     \begin{subfigure}[b]{\textwidth}
         \centering
         \includegraphics[width=0.6\textwidth]{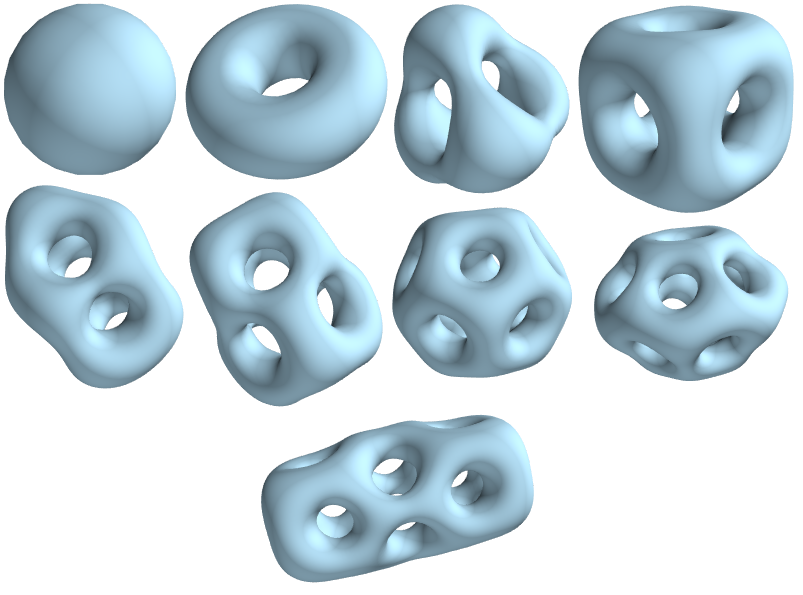}
         \caption{Skyrmions with baryon number from 1 to 8, including the two energy-degenerate solutions with different symmetries, $8a$ (second row) and $8b$ (third row). }
         \label{fig:1_8}
     \end{subfigure}
     
     \begin{subfigure}[b]{\textwidth}
         \centering
        \includegraphics[width=0.6\textwidth]{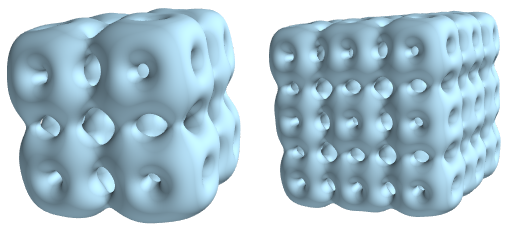}
         \caption{ The $B=32$ and $B=108$ Skyrmions. They present a manifest cubic symmetry, as they can be thought of being composed of individual $B=4$ ``bricks''. }
         \label{fig:32_108}
     \end{subfigure}
     \caption{Energy density iso-surfaces of the classical Skyrmion configurations used in this paper}
\end{figure}

In order to compute the gravitational form factors of light nuclei, we start by generating the static energy-minimizing classical field configuration corresponding to each of the topological sectors. We do this by first constructing a numerical ansatz using the  rational map approximation \cite{Manko:2007pr}, and allow this initial configuration to relax to the true minimizer via gradient descent.
We remark that this is an increasingly difficult task for values of the topological charge $B\gtrsim 8$, not only due to the increase of the solution's mean radius, which requires a numerical simulation on a lattice with an increasing number of points, but also because of the flatness of the field configuration landscape, which in general presents multiple local minima with very similar energies but different symmetries and shapes, as recently shown by Gudnason and Halcrow \cite{Gudnason:2022jkn}. This fact points towards the failure of the rigid quantization approximation---in which quantum effects do not modify the classical shape of the Skyrmions---for higher baryon charges in general. We have, nevertheless, constructed also solutions with large baryon charge and cubic symmetry, namely, the $B=32=4\times 2^3$ and $B=108=4\times 3^3$ Skyrmions, which are expected to not be affected by spin nor isospin quantum effects in their ground state, and are considered to be the minimum energy configurations due to their high degree of symmetry. These solutions are illustrated in \cref{fig:1_8} ($1\le B\le 8$) and \cref{fig:32_108} ($B=32,108$) where energy density iso-surfaces are plotted.  For $B=8$, we consider two energy-degenerate solutions\footnote{ The difference in energies of both solutions is less than $1\%$. The precise number (and even its sign) will depend on the parameter values.} with different symmetries. 

\begin{figure}
   \centering
    \includegraphics[scale=0.9]{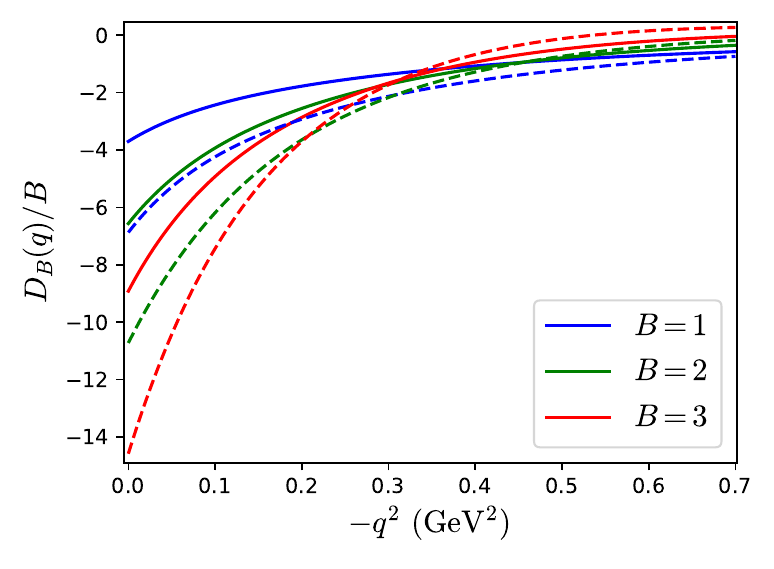}
    \caption{ The $D$ gravitational form factor of the Skyrmions with $B=1,2,3$, normalized by $B$, for $\lambda=0$ (solid) and $\lambda^2=3$ $\Mev\fm ^3$ (dashed).}
  \label{123}
\end{figure}

After obtaining the classical solutions, we  used the formula \eqref{main} (with $M_N$ replaced by the respective nuclear masses) to compute the $D$-form factor for the first eight topological sectors, for which the ground state is well known, as well as for the $B=32$ and $B=108$ cubic Skyrmions.
The results for the first three Skyrmions  are shown in \cref{123}.  We have used the same values for the parameters in the Lagrangian as in \cite{Cebulla:2007ei}. Thus the $B=1$ result is in agreement with \cite{Cebulla:2007ei}, while the $B=2,3$ results are new.  
In the same plot, we also show the results obtained with a second set of parameters which includes a nonzero value of the sextic coupling constant $\lambda^2= 3 \Mev\fm ^3$. This choice is motivated by some previous studies on the symmetry energy of infinite nuclear matter within the same model \cite{Adam:2022aes}, as well as the EOS of neutron stars \cite{Adam:2022cbs}.  Finding classical solutions with nonzero $\lambda$ via gradient-descent based algorithms becomes numerically more challenging for arbitrarily large values of this parameter. With our choice of $\lambda^2= 3 \Mev\fm ^3$, we have been able to obtain trustable solutions only up to $B=32$. We find that the main effect of the sextic interaction is to increase the magnitude (in absolute value) of the D-term $D(0)$.  This can be traced back to the fact that the sextic term represents a repulsive interaction, which makes the size of the soliton to grow to pick up more contributions from the large-radius region in the integral (\ref{ss}). However, apparently it contradicts with the recent observation \cite{Fujita:2022jus} in the Sakai-Sugimoto model \cite{Sakai:2004cn} that the repulsive interaction due to the omega meson exchange decreases the magnitude of the D-term. Unlike in \cite{Fujita:2022jus}, in the present model the omega meson is treated as a static field represented by  the sextic coupling \cite{Jackson:1985yz}, and induces only a diagonal term $T_{ij}\propto c_6\delta_{ij}$ in the energy momentum tensor (\ref{classical}) which does not directly contribute to the $D$-form factor.  Thus the value of the D-term seems to be rather sensitive to  different   (static or dynamical) treatments for the omega meson field. 
 With the choice $\lambda^2=3$ MeV fm$^3$, our result $D(0)\approx -7$ for $B=1$ is on the higher end in magnitude among model results in the literature, see, e.g., \cite{Goeke:2007fp,Cebulla:2007ei,Neubelt:2019sou,Fujita:2022jus,Won:2022cyy,Chakrabarti:2020kdc}.


\begin{figure}
    \centering
    \includegraphics[scale=0.55]{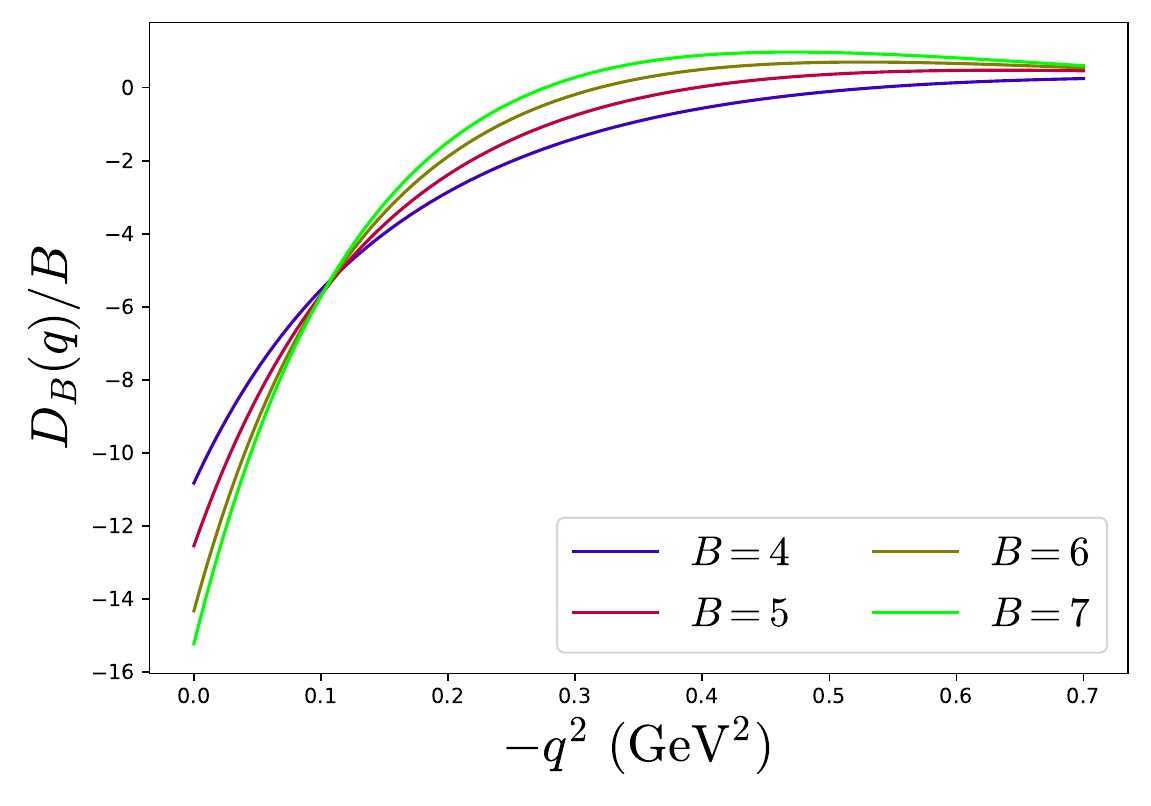}
    \caption{The $D$ gravitational form factor of the Skyrmions with $B=4,5,6$ and $7$, normalized by $B$, for the $\lambda=0$ case.}
    \label{subfig:mediumB_D}
\end{figure}

\begin{figure}
    \centering
    \includegraphics[scale=0.55]{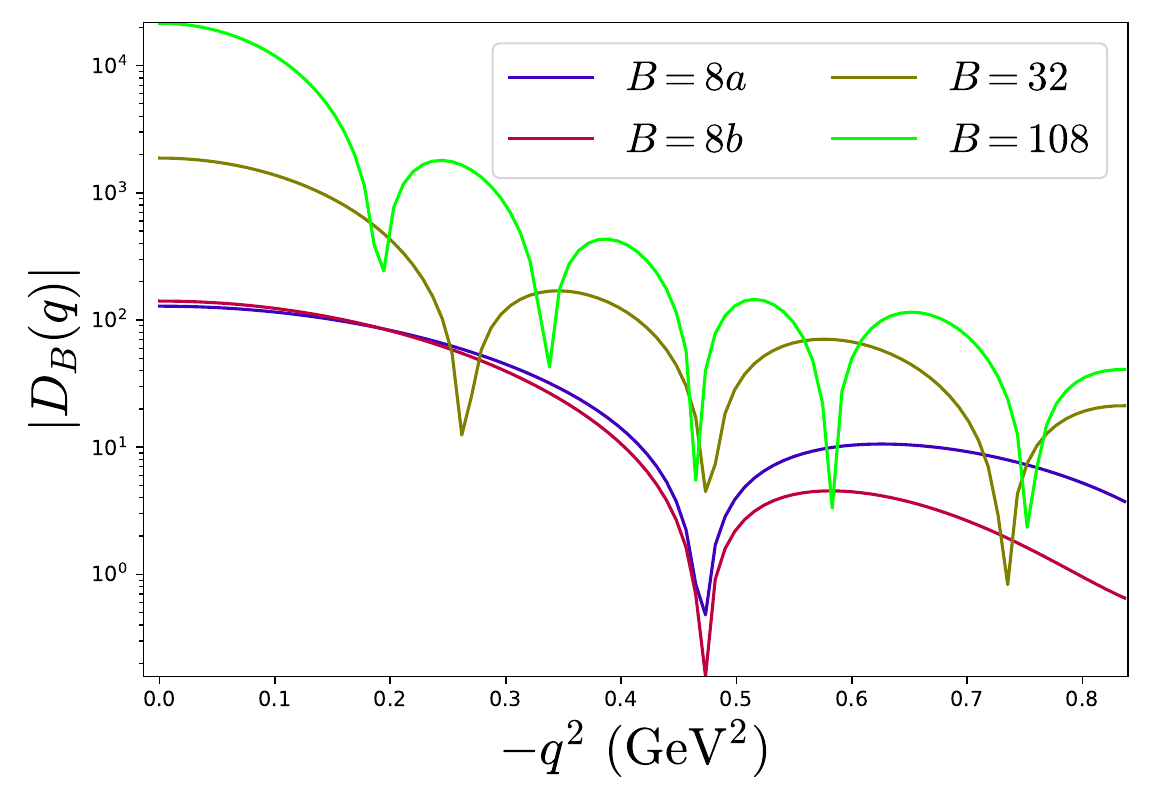}
    \caption{The  $D$ gravitational form factor (absolute value, not normalized by $B$) of the Skyrmions with $B=8,32,108$, also for $\lambda=0$. Cusps mean that the form factor flips signs and  oscillates around zero. }
    \label{subfig:highB_D}
\end{figure}

\begin{table}
\begin{tabular}{|c|c|c|c|c|c|c|c|c|c|c|c|}
\hline
$B$ & 1 & 2 & 3 & 4 & 5 & 6 & 7 & 8$a$ & 8$b$ & 32 & 108 \\
\hline 
$D(0)$ & -3.701 & -13.126 & -26.757 & -43.304 & -62.72 & -85.95 & -106.596 & -128.368 & -140.816 & -1.874$\times10^3$ & -2.152$\times10^4$ \\
\hline
\end{tabular}
\caption{The values of the D-term  $D(t=0)$ at $\lambda=0$.} 
\end{table}

\begin{figure}
         \centering
         \includegraphics[scale=0.9]{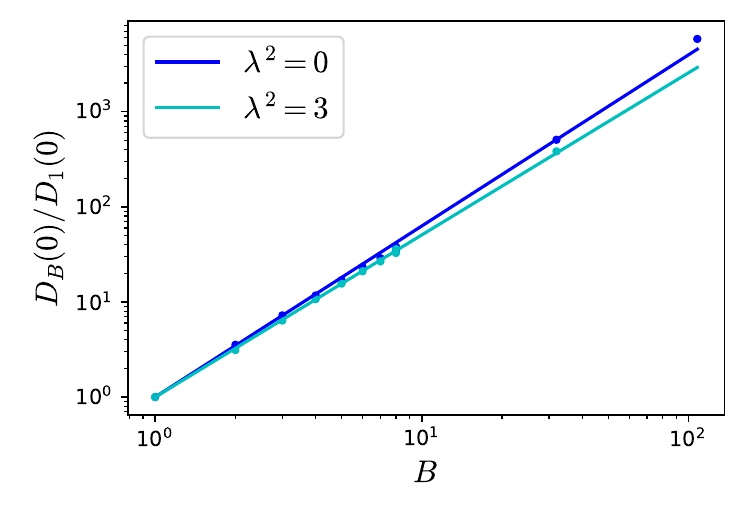}
         \caption{Dependence of the $D$ term on the baryon charge $B$ for Skyrmions with $B\le 8$, $B=32$ and $B=108$. The cases  $\lambda^2=0$ (dark blue) and $\lambda^2=3$ $\Mev\fm ^3$  (light blue)  are shown.}
         \label{subfig:DB_D0}
    \label{fig:D1234}
    
\end{figure}

The results for the $B>3$ solutions with $\lambda=0$ are plotted in  \cref{subfig:mediumB_D} and \cref{subfig:highB_D},  and the values at the origin $D(0)$ are summarized in Table 1.
All the curves start with a negative value, with their absolute value increasing with $B$, and cross zero at least once for some value of $|t|$. The number of nodes increases with $B$, and they appear earlier for higher values of $B$. That   $D(t=0)$ is negative is consistent with the general expectation  \cite{Polyakov:2018zvc}, but the sign of $D(t)$ for $t\neq 0$ is not constrained by any physical argument. Technically, the oscillation is caused by the spherical Bessel function $j_2$ in (\ref{main}) convoluted  with the increasingly flatter density profile of larger-$B$ solutions. 

The particular values of the $D_B(0)$ for each topological sector will depend strongly on the specific values of the parameters of the model. However, their  scaling with the topological charge, or atomic number, of the nuclei is a genuine prediction of the Skyrme model.
We have found that the $B$-dependence is very well fitted by a simple power-law $D_B(0)\propto B^\beta$, or equivalently, 
\begin{equation}
   \beta\equiv \frac{\log\frac{D_B(0)}{D_1(0)}}{\log B},
   \label{eq:phenolinear}
\end{equation}
is a constant 
as demonstrated in \cref{subfig:DB_D0}.
The value of $\beta$ is found to be $\beta\approx1.8$ for $\lambda^2=0$ and $\beta\approx 1.7$ for $\lambda^2=3  \Mev\fm ^3$. 
In comparison, we note that the liquid drop model of nuclei predicts $\beta=7/3$ \cite{Polyakov:2002yz} which is consistent with the result in the Walecka model  $\beta\approx 2.26$ \cite{Guzey:2005ba}. On the other hand, a microscopic approach using the nonrelativisitic  nuclear spectral function predicts  $\beta=1$ \cite{Liuti:2005qj}. 

\begin{figure}
    \centering
    \includegraphics[scale=0.6]{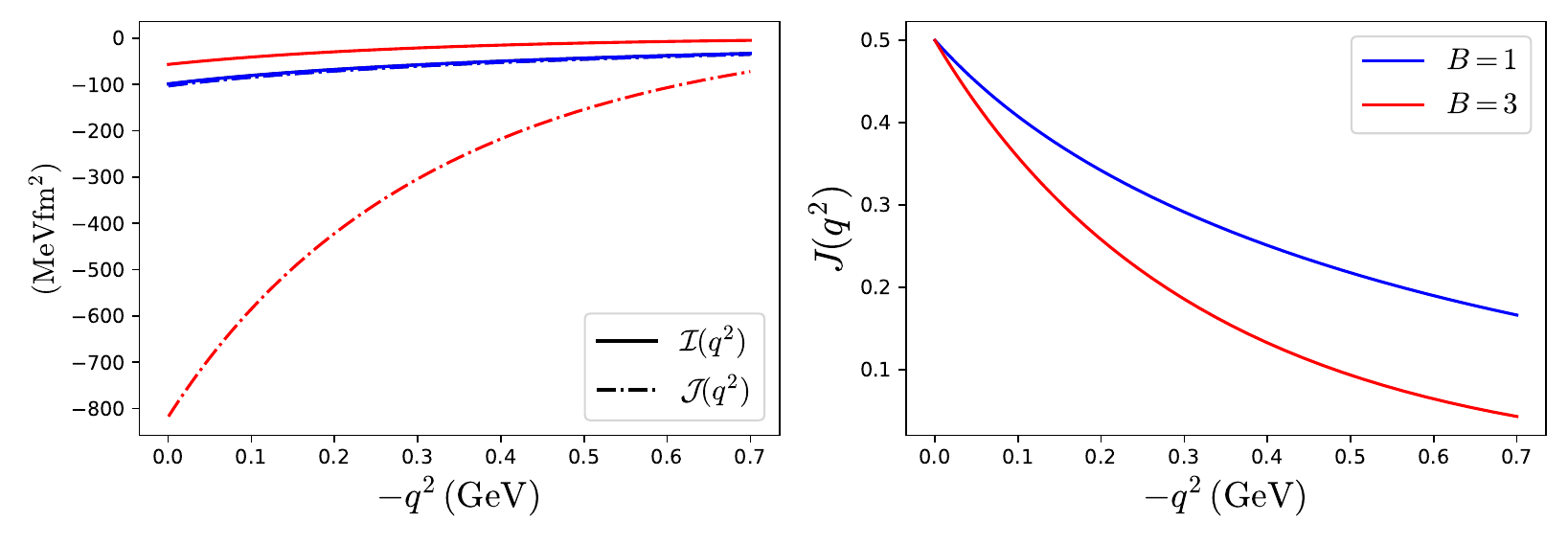}
    \caption{Left: Values of the functions $\mathcal{I}(q)$ and $\mathcal{J}(q)$ for the $B=1$ and $B=3$ cases. Right: The gravitational form factor $J(q^2)$ for the same nuclei.}
    \label{fig:IJcurves}
\end{figure}

Finally, in \cref{fig:IJcurves} we show the results for the angular momentum form factor $J(q^2)$ together with the isospin ${\cal I}$ and rotation 
 ${\cal J}$ currents defined in (\ref{only}) for the $B=1,3$ solutions.   For $B=1$, ${\cal I}$ and ${\cal J}$ are equal due to the symmetry of the hedgehog configuration. The total $J$-form factor reproduces the result in   \cite{Cebulla:2007ei}.  See, e.g.,  \cite{Goeke:2007fp,Neubelt:2019sou,Won:2022cyy} for calculations in other models.
For $B=3$, we see that the angular momentum distribution is more localized at small momentum transfer, or at larger radius in position space. 

\section{Conclusions}

In this paper, we have calculated one of the gravitational form factors, the $D(t)$-form factor,  for nuclei with baryon numbers  $B=1,2,3,4,5,6,7,8,32,108$ in the Skyrme model. While the electromagnetic and axial form factors have been previously computed for light nuclei  in this model \cite{Braaten:1988bn,Carson:1991fu}, its application to nuclear GFFs is new. Despite different patterns of nuclear deformation, the monopole $D(t)$-form factor is given by the single formula (\ref{main}) which can be readily evaluated for a given classical configuration.  Our approach can be generalized to other GFFs and we have presented a concrete  example, the angular momentum form factor $J(t)$, for the $B=3$ solution. 

There are a number of future directions.  In this work we neglected the ${\cal O}(J^2,I^2)$ terms in (\ref{quant}) invoking the large-$N_c$ approximation. However, these quantum corrections are necessary to classify different isobars and study the detailed spin and isospin effects for individual nuclei and their excited states. For the $B=1$ solution, they have been included in   \cite{Kim:2020lrs} to compute the GFFs of the $\Delta$-resonances. Additional complications arise for $B>1$ solutions due to their non-sphericity, and it remains to be seen whether group theory techniques help to mitigate the problem.  
We also neglected the spin-dependent quadrupole gravitational form factors for the deuteron. In future work, they can be  evaluated and extended to other spin-1 nuclei.  Moreover, we have not explored solutions with $B>8$ except for the rather special $B=32,108$ solutions which are realized as cubic `$\alpha$-clusters'.  Numerical solutions are available in the literature  \cite{Feist:2012ps} and they can be used to study the $B$-dependence of GFFs  more closely. An important caveat when considering large Skyrmions is the failure of the rigid body quantization. A more involved quantization procedure, including some vibrational degrees of freedom and  allowing for deformations of the classical Skyrmion shape has been developed during the last decade with relatively successful applications to light nuclei, such as the $B=5$ \cite{Gudnason:2018aej}, $B=7$ \cite{Halcrow:2015rvz} or $B=16$ \cite{Halcrow:2019myn} solutions. The study of the GFF of larger Skyrmions within a vibrational quantization is also a straightforward extension of our present work, which nevertheless falls outside our original scope, namely, to determine the dominant contribution to the $D$ term of each Skyrmion. 
Last but not least, recently in \cite{Fujita:2022jus}, the nucleon D-term has been studied in holographic QCD in the Sakai-Sugimoto model \cite{Sakai:2004cn}. It has been observed that, in order to properly compute GFFs at finite $t\neq 0$, one  must include glueballs. While it is known that the Skyrme model can be derived from the Sakai-Sugimoto model, glueballs are usually lost in this reduction. It appears to be a challenging task to consistently restore these degrees of freedom and study their impact on GFFs.

\section*{Acknowledgements}

We thank Shigeki Sugimoto for discussions, and Chris Halcrow for useful comments.
A.~G.~M. thanks Brookhaven National Laboratory, where this work was initiated, for hospitality. 
Y.~H. is supported by the U.S. Department of Energy under Contract No. DE-SC0012704, and also by  Laboratory Directed Research and Development (LDRD) funds from Brookhaven Science Associates. A.~G.~M. and M.~H. acknowledge financial support from the Ministry of Education, Culture, and Sports of Spain (Grant No. PID2020-119632GB-I00), the Xunta de Galicia (Grant No. INCITE09.296.035PR, and the European Union ERDF.
 A.~G.~M. is grateful to the Spanish Ministry of Science, Innovation and Universities, and the European Social Fund for the funding of his predoctoral research activity. M.~H. thanks the Xunta de Galicia (Consellería de Cultura, Educación y Universidad) for the funding of his predoctoral activity through \emph{Programa de ayudas a la etapa predoctoral} 2021.


\nocite{BjarkeGudnason:2018bju}
\bibliography{Bibliography}

\end{document}